\newif\ifColor\Colortrue \Colorfalse
\Colortrue
\PassOptionsToPackage{super,comma,numbers,sort&compress}{natbib}  
\documentclass[%
reprint,
superscriptaddress,
 amsmath,amssymb,
 aps,
]{revtex4-2}

\usepackage{lineno}
\usepackage{graphicx}
\usepackage[utf8]{inputenc}
\usepackage{amssymb}
\usepackage{amsmath}
\usepackage{amsfonts}
\usepackage{upgreek}
\usepackage{gensymb}
\usepackage{setspace}
\usepackage{booktabs}
\usepackage[english]{babel}
\usepackage{mathtools}
\usepackage{mhchem}[version=4]
\usepackage{hyperref}
\usepackage[final]{microtype}
\usepackage{csquotes,lmodern}
\usepackage{etoolbox}
\usepackage{color}
\usepackage{multibib}
\usepackage{siunitx}
\usepackage[normalem]{ulem}
\usepackage{dcolumn}
\usepackage{bm}

\begin{document}

%\title{Improved thermoelectric conversion efficiency of \ce{Fe2VAl}-based thermoelectric modules}
\title{High-entropy \ce{Fe2VAl}-based thermoelectric modules with improved conversion efficiency}

\author{M. Parzer*}
\affiliation{Institute of Solid State Physics, TU Wien, 1040 Vienna, Austria}
%\email{Michael_parzer@yahoo.de}
\author{G. Roy*}
\affiliation{Institute of Mechanics, Materials, and Civil Engineering, IMAP, UCLouvain, 1348 Louvain-la-Neuve, Belgium}
%\email{geoffrey.roy@uclouvain.be}
\author{F. Garmroudi}
\affiliation{Materials Physics Applications -- Quantum, Los Alamos National Laboratory, Los Alamos, 87545 New Mexico, USA}
\author{P. Ziolkowski}
\affiliation{German Aerospace Center (DLR) – Institute for Frontier Materials on Earth and in Space, Linder Höhe, D-51147 Cologne, Germany}
\author{T. Konegger}
\affiliation{Institute of Chemical Technologies and Analytics, TU Wien, 1060 Vienna, Austria}
\author{E. Bauer}
\affiliation{Institute of Solid State Physics, TU Wien, 1040 Vienna, Austria}
\author{P.J. Jacques}
\affiliation{Institute of Mechanics, Materials, and Civil Engineering, IMAP, UCLouvain, 1348 Louvain-la-Neuve, Belgium}

\begin{abstract}
Thermoelectric (TE) materials enable the direct conversion of heat into electricity and are attractive for sustainable energy applications. For practical deployment, TE materials must combine high efficiency with low cost, non-toxicity, and scalability. In this work, we optimize the TE performance of low-cost and robust \ce{Fe2VAl}-based full-Heusler compounds through high-entropy engineering: a synergistic combination of heavy-element doping and controlled off-stoichiometry results in substitutional disorder on all lattice sites, triggering one of the lowest lattice thermal conductivities, $\kappa_\text{L}\sim2.3$\,W\,m$^{-1}$\,K$^{-1}$, reported so far for full-Heusler systems. The resulting materials exhibit improved values of the average figure of merit $zT_\text{ave}\approx 0.3$ from 300--500\,K.
To demonstrate reproducibility and technological relevance,  
a full TE module (TEM) based on the optimized alloys was fabricated and characterized. Scaled-up material batches were synthesized by hot pressing, exhibiting TE properties in excellent agreement with laboratory-scale samples, with only slightly increased resistivities in absence of post-annealing treatments. Owing to the excellent mechanical workability of \ce{Fe2VAl}-based materials, the TEM legs were directly brazed onto copper electrodes, enabling robust module fabrication.
A (6\,$\times$\,6)-leg TEM was assembled and systematically characterized. The device exhibits the highest output power and one of the highest conversion efficiencies reported to date for \ce{Fe2VAl}-based generators over the broad temperature range of 300--673\,K, underscoring the potential of this material system for scalable TE energy harvesting.

%\vspace{0.5cm}
%\textcolor{blue}{Remark by E.B.: while ``high entropy'' alloying sounds very elegant as title of our manuscript, I am not quite sure, whether we really have the classical recipe for a standard high entropy material. Although we do have 5 elements, they are not in near-equal (or equal) proportions and we do not have thermodynamic infos  from programs like CALPHAD.  If, however, you find it better to keep this term, then we can wait for the referee's comments.  }
\end{abstract}

\maketitle

\onecolumngrid
\twocolumngrid
\section{Introduction}
Thermoelectric (TE) devices exploit the Seebeck effect to convert temperature gradients directly into electrical power within a single solid‐state element, obviating moving parts and offering unparalleled reliability. This simplicity underpins the appeal of TE generators (TEGs) for waste‐heat recovery across a range of applications \cite{hendricks2022keynote,zhang2022micro,pecunia2023roadmap}. Over the past two decades, intensive materials research has substantially improved the intrinsic materials figure of merit
\begin{equation}
zT=\frac{S^2\sigma}{\kappa}T\;,
\end{equation}
where $S$ is the Seebeck coefficient, $\sigma$ the electrical conductivity, and $\kappa$ the thermal conductivity—yielding peak $zT$ values approaching 3 in some compounds \citep{yan2022high, ryu2023best}. In contrast, most thermoelectric devices in current commercial use rely on \ce{Bi2Te3}-based alloys with typical figures of merit of $zT \approx 1$, a material system developed in the 1960s and still manufactured using largely unchanged processing routes \citep{witting2019thermoelectric}. Similarly, space-grade TEGs continue to employ SiGe-based compounds, selected less for maximal efficiency than for their exceptional stability and long-term reliability under extreme operating conditions \citep{aswal2016key}.

These examples illustrate that attaining a high $zT$ is only one facet of material suitability. For TE technology to achieve broad commercialization, considerations of chemical and mechanical stability, element abundance, manufacturability, cost, and sustainability become equally critical \citep{yazawa2011cost,yan2022high,bos20252025}. In particular, secondary properties—such as ductility, machinability, contactability, resistance to oxidation and compatibility with large‐scale device assembly—are frequently neglected in materials screening but often constitute the principal barriers to deploying new thermoelectrics in functional modules \citep{yan2022high}. Addressing these challenges through post‐synthesis engineering or complex fabrication protocols adds time, expense, and risk, and in some cases may prove intractable.

In this work, we pursue a complementary strategy: instead of chasing record $zT$ alone, we optimize the thermoelectric performance of \ce{Fe2VAl}-based full‐Heusler compounds that inherently combine favorable secondary properties with still sizeable $zT$ near room temperature. As showcased in previous studies, the material can be directly brazed onto copper plates, forgoing any protection layers or anti-oxidation coatings \citep{roy2019global}. Moreover, previous research on \ce{Fe2VAl}-based TE modules has proven the good stability in long-term tests and direct application \citep{mikami2009development, mikami2011power, mikami2014evaluation}. Recently, also thin-film TE modules were fabricated utilizing \ce{Fe2VAl}-based materials, which enable microwatt power generation for sensor applications with very small TEGs \citep{bourgault2023unlocking}.
%Here, we demonstrate further enhancements in $zT$ of \ce{Fe2VAl}, using high-entropy alloying, while retaining excellent material workability. Using large-scale materials synthesis, we build up a fully working 6\,x\,6-leg TE module from non-nanostructured bulk materials and fully characterize its TE performance, yielding one of the highest efficiencies $\eta$ measured for \ce{Fe2VAl}-based modules.
Here, we demonstrate further enhancements in $zT$ of \ce{Fe2VAl}, using high-entropy alloying, a strategy that has proven successful in other TE material systems \citep{jiang2021high, jiang2022high, ghosh2024high}.  Moreover, we validate the retention of excellent material workability by using large-scale materials synthesis. We build up a fully working 6\,x\,6-leg TE module from non-nanostructured bulk materials and fully characterize its TE performance, yielding one of the highest efficiencies $\eta$ measured for \ce{Fe2VAl}-based modules.

\begin{figure*}
\centering
\includegraphics[width=1\textwidth]{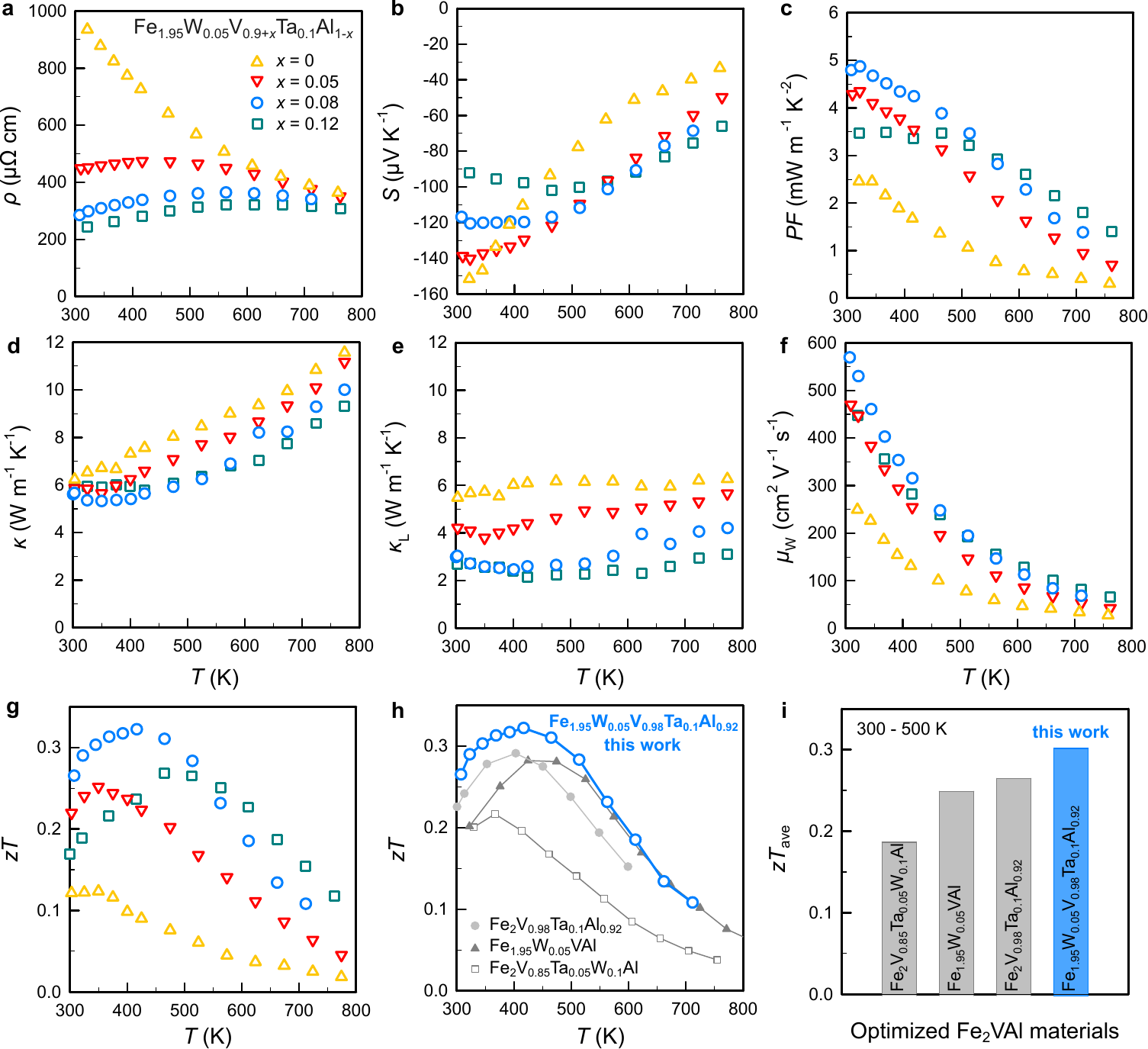}
\caption{\textbf{Optimization of \ce{Fe2VAl}-based materials via heavy-element co-substitution and off-stoichiometry.} \textbf{a}, Temperature-dependent electrical resistivity; \textbf{b}, Seebeck coefficient; \textbf{c}, power factor ;\textbf{d}, thermal conductivity; \textbf{e}, lattice thermal conductivity; \textbf{f}, weighted mobility; and \textbf{g}, dimensionless figure of merit of \ce{Fe_{1.95}W_{0.05}V_{0.9+x}Ta_{0.1}Al_{1-x}} full-Heusler compounds; \textbf{h}, Dimensionless figure of merit of best-performing sample from this work compared to previously optimized full-Heusler materials taken from \citep{takagiwa2023effect, fukuta2022improving,garmroudi2025recent}; \textbf{i} Average $zT$ from 300 to 500\,K, outperforming all other \ce{Fe2VAl}-based compositions without microstructure refinement.}
\label{fig:properties}
\end{figure*}

\section{Optimization of \ce{Fe2VAl}-based materials}
\ce{Fe2VAl}-based \citep{nishino1997semiconductorlike,nishino2006thermal,lue2007thermoelectric,garmroudi2021boosting} and recently also \ce{Ru2TiSi}-based \cite{fujimoto2023enhanced,garmroudi2025thermoelectric,garmroudi2025recent,garmroudi2026orbital} Heusler compounds with 24 valence electrons have drawn considerable attention for both fundamental studies \cite{nishino1997semiconductorlike,singh1998electronic,weht1998excitonic,parzer2025mapping} and thermoelectric (TE) applications \cite{mikami2009development,bourgault2023unlocking,dominguez2025thermoelectric,tarachandhigh,jha2025high} due to their peculiar electronic structure, marked by steep densities of states near the Fermi level ($E_\text{F}$). Although the exact width of the band gap of \ce{Fe2VAl} remains debated, the system is typically described as a narrow-gap semiconductor or semimetal with a quasi-low-dimensional Fermi surface for the Fe $e_g$ conduction band states \citep{bilc2015low, garmroudi2022large}.
 These features enable large Seebeck coefficients even at relatively high carrier concentrations. As a result, $n$-type \ce{Fe2VAl}-based materials exhibit exceptional power factors ($PF = \sigma S^2$), surpassing those of \ce{Bi2Te3} and other state-of-the-art semiconductors, potentially reaching up to 10\,mW\,m$^{-1}$K$^{-2}$ for $n$-type materials and  up to $4$\,mW\,m$^{-1}$K$^{-2}$ for $p$-type materials \citep{garmroudi2021boosting,miyazaki2013thermoelectric}.

Nonetheless, their simple crystal structure and high Debye temperature result in inherently high lattice thermal conductivities ($\kappa_\text{L}$), which severely limit the figure of merit $zT$ (cf. Eq. 1). Consequently, much of the research has focused on strategies to reduce $\kappa_\text{L}$ without compromising the favorable electronic transport properties. One key approach involves tuning the electronic structure via off-stoichiometry and targeted substitution. A broad range of off-stoichiometry schemes has been investigated—-including Fe/V \citep{lue2002thermoelectric, nishino2014doping, takagiwa2023effect}, V/Al \citep{miyazaki2013thermoelectric, alleno2023optimization, asai2025tailoring}, Fe/Al \citep{soda2014semiconducting, diack2022influence} and even Fe/V/Al \citep{parzer2022high, parzer2024semiconducting}--to control carrier concentration and optimize $PF$, sometimes even via exotic mechanisms such as carriers interacting favorably with the magnetic degrees of freedom \cite{tsujii2019observation,tsujii2024effect}, while also reducing the thermal conductivity via disorder and aliovalent doping-induced lattice softening. 

\begin{figure}
\centering
\includegraphics[width=0.45\textwidth]{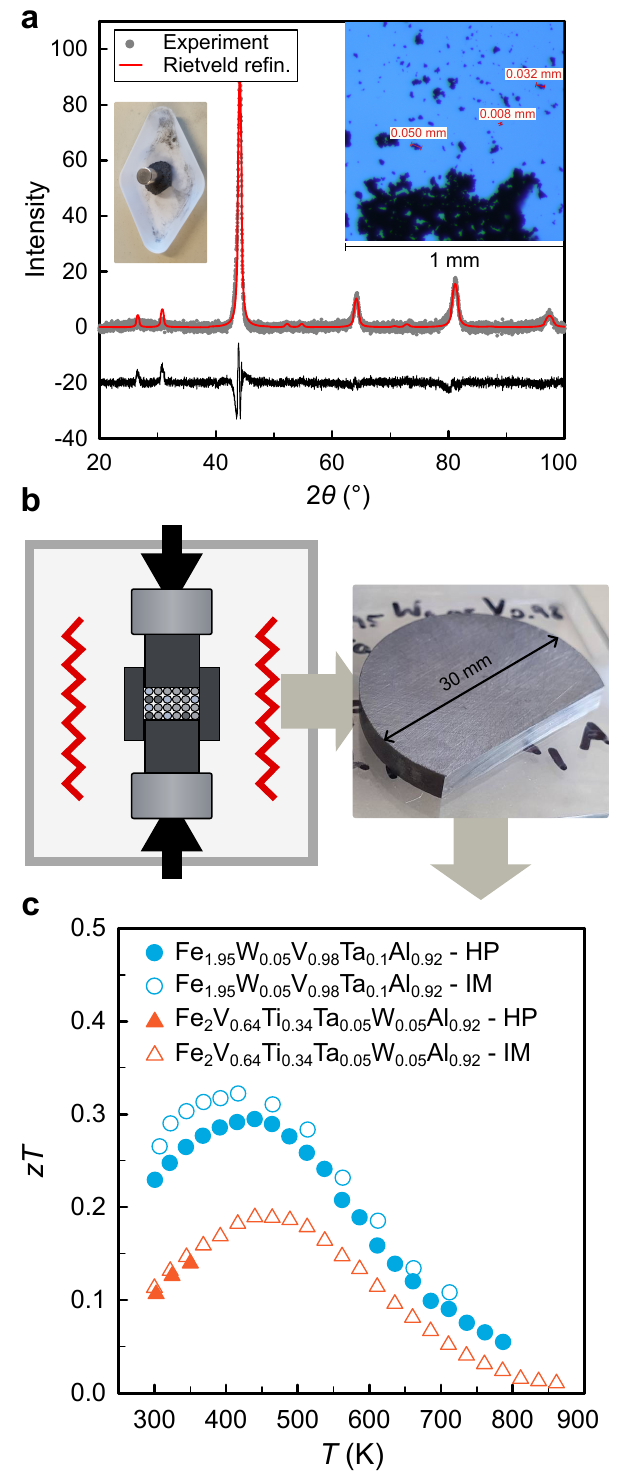}
\caption{\textbf{Powder analysis and hot-press synthesis of optimized \ce{Fe2VAl}-based thermoelectrics.} \textbf{a}, X-ray diffraction pattern and Rietveld refinement of ball-milled powder. Inset shows an image of the product taken with an optical microscope. \textbf{b}, Schematic of the consolidation of the powder via hot press and image of obtained pellets. \textbf{c}, Dimensionless figure of merit of induction-melted (IM) samples compared to samples cut from the hot-pressed (HP) pellets.}
\label{fig:hot-press}
\end{figure}
To further suppress $\kappa_\text{L}$, alloying with elements of significantly different mass or atomic size has proven effective \citep{terazawa2012effects, hinterleitner2020stoichiometric,garmroudi2021solubility}. The resulting mass and strain field fluctuations scatter heat-carrying phonons, especially at high frequencies \citep{kimura2020local}. Recently, Fukuta et al. achieved $zT$ values up to 0.29 at 400 K by combining Ta substitution at the V site with Al/V off-stoichiometry in \ce{Fe2V_{0.9+x}Ta_{0.1}Al_{1-x}} with $x=0.1$ \citep{fukuta2022improving}. Similarly, Takagiwa and Iwasaki combined heavy 5$d$ element substitution and off-stoichiometry in nominally Fe/W substituted \ce{Fe_{2-x}W_{x}VAl}. Since W preferentially occupies the V site, this substitution results in V/W substitution together with Fe/V antisite defects. The authors achieved a high $zT=0.28$ at 423 K for $x=0.05$ and 0.1 \citep{takagiwa2023effect}.

Here, building on a combined "high-entropy" approach of targeted off-stoichiometry and heavy-element doping \cite{parzer2025enhanced}, we identify promising compositions based on \ce{Fe_{1.95}W_{0.05}V_{0.9+x}Ta_{0.1}Al_{1-x}} with substitution and defects on all crystallographic lattice sites. Thereby, we are able to optimize both electronic and thermal transport achieving a high $zT$ up to 0.32 at 417 K. To translate these insights into practical improvements, most promising compositions were synthesized on a lab scale and systematically characterized. The ultimate goal was to develop efficient $p$- and $n$-type \ce{Fe2VAl}-based materials for integration into thermoelectric modules, the performance of which will be discussed in \autoref{TEM}.

Figure \ref{fig:properties} summarizes the thermoelectric properties and performance of $n$-type \ce{Fe_{1.95}W_{0.05}V_{0.9+x}Ta_{0.1}Al_{1-x}} thermoelectrics synthesized by induction melting. The self-substitution Al/V controls the carrier concentration and simultaneously introduces strong disorder scattering for phonons resulting in a low lattice thermal conductivity $\kappa_\text{L}\approx 2.3$\,W\,m$^{-1}$\,K$^{-1}$ for $x=0.12$. To the best of our knowledge, this constitutes the lowest $\kappa_\text{L}$ for \ce{Fe2VAl}-based full-Heusler compounds without considering nano-structuring or thin film deposition. At the same time, the weighted carrier mobility $\mu_\text{W}$ calculated via the formula from ref.\,\cite{snyder2020weighted} retains exceptionally high values $\mu_\text{W}=400-600$\,cm$^2$\,V$^{-1}$\,s$^{-1}$, whereas nano-structuring such as reducing grain size via ball milling or severe plastic deformation can often lead to reduced $\mu_\text{W}$ in \ce{Fe2VAl}-based materials \cite{garmroudi2025decoupled}. Interestingly, $\mu_\text{W}$ even increases with doping initially, which we attribute to a decrease of bipolar conduction as the Fermi level is shifted further into the conduction band. Indeed, the Seebeck coefficient in Fig.\,\ref{fig:properties}b showcases that the maximum $S_\text{max}$ shifts to higher temperatures with increasing $x$ as does the onset of bipolar thermal transport seen as a temperature-dependent increase in the lattice thermal conductivity (Fig.1c). The optimal $zT$ is achieved for $x=0.08$, surpassing that of other W and Ta co-doped systems, where only one or two lattice sites were substituted \cite{fukuta2022improving,takagiwa2023effect,garmroudi2025recent}, in the entire measured temperature range (see Fig.\,\ref{fig:properties}h). Figure \ref{fig:properties}i shows that, between 300 and 500\,K, the average dimensionless figure of merit $zT_\text{ave}$ achieved here ranks among the largest values for \ce{Fe2VAl}-based full-Heusler thermoelectrics.

\ce{Fe2VAl}-based $p$-type materials were optimized using a similar high-entropy approach, resulting in an optimal $zT$ of 0.2 between 450--500\,K (see Fig. 2c) for a sample with the composition \ce{Fe2V_{0.64}Ti_{0.34}Ta_{0.05}W_{0.05}Al_{0.92}}. The temperature-dependent thermoelectric properties of the $p$-type material are summarized in the Supporting Information (Fig. S1).

\section{Large scale synthesis via hot press}
To achieve larger sample batches and enable upscaling of the TE material production, we employed hot pressing of pre-synthesized TE material powder. To this aim, 30\,g of pre-synthesized material powder were fabricated using compositions displaying the highest TE efficiency, namely \ce{Fe_{1.95}W_{0.05}_V_{0.98}Ta_{0.1}Al_{0.92}} as $n$-type and \ce{Fe2V_{0.64}Ti_{0.34}Ta_{0.05}W_{0.05}Al_{0.92}} as $p$-type. The material was milled into a powder utilizing vibrational ball milling as described in \autoref{methods}. As showcased in the top-right inset of \autoref{fig:hot-press}\,a, this resulted in average grain sizes of around $d\approx50\,\si{\mu m}$. Notably, the powder exhibited strong magnetic attraction as highlighted qualitatively by the photo of the powder sticking to a neodymium permanent magnet in the left inset of Fig.\,2a. This is also hinted at by the tendency of grain clumping during light microscopy measurements. X-ray diffraction (XRD) measurements (see \autoref{fig:hot-press}\,a), conducted on the ball-milled powder, revealed complete A2-type disorder in the crystal structure, which was induced during the milling process and is a well known consequence of severe mechanical stress in \ce{Fe2VAl}-based Heusler compounds \citep{graf2011simple,maier2016order}. The fully disordered A2-type structure is characterized by the missing L2$_1$ symmetry (111) and (200) symmetry peaks at 27\,$^\circ$ and 31\,$^\circ$, respectively. Additionally, recent studies have revealed that such antisite disorder leads to strong magnetic moments in \ce{Fe2VAl}-based samples \citep{alleno2020structure,garmroudi2022anderson,garmroudi2023unveiling}, bringing the qualitative observations mentioned above in good agreement with the XRD results.

\begin{figure}[b]
\centering
\includegraphics[width=0.45\textwidth]{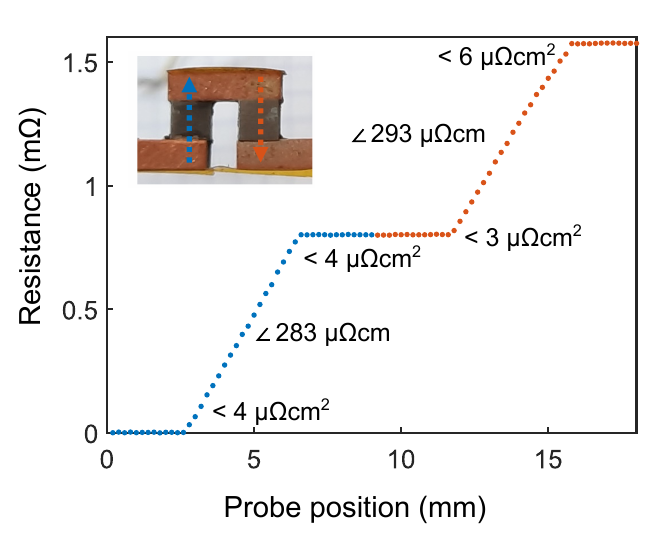}
\caption{\textbf{Resistance scan on a 2-leg module}. The plot depicts the total resistance over the probe position measured using a custom-built voltage scanning probe. The scanning step colored in blue represents the scan of the $n$-leg, the second step (colored in red) represents the scanning of the $p$-type leg.}
\label{fig:rScan}
\end{figure}
Subsequently, from each sample, an appropriate amount of powder was weighed in to achieve hot-pressed disks of 4\,mm height and 30\,mm diameter (see \autoref{fig:hot-press}\,b on the right. The powders were uniaxially hot-pressed at 1100\,$^\circ${}C in Ar atmosphere (0.1\,MPa) at a compaction pressure of 35\,MPa (FCT HP-W 150/200-2200, Germany), as sketched in \autoref{fig:hot-press}\,b. After filling the powders into the graphite tools (inner diameter 30\,mm) and vacuum-purging the system, the samples were pre-compacted at 15\,MPa. The temperature was increased to 650\,$^\circ${}C, where the final pressure of 35\,MPa was applied, before further heating to the final temperature of 1100\,$^\circ${}C (holding time: 1\,h). After cooling, the pressure was released, the samples were demolded, and the samples surface was cleaned from residual graphite foil used as separating agent. During the whole process, heating and cooling rates of 10\,K.min$^{-1}$ were used.

From each of the discs, pieces for sample characterization were cut out and measured. \autoref{fig:hot-press}\,c shows the comparison of the temperature-dependent $zT$ between the small induction-melted $n$- and $p$-type samples with the hot-pressed sample pieces, respectively.
%The individual resistivity $\rho(T)$, $S(T)$ and $\lambda(T)$ are shown in Fig. S? in the supplemental material.
Clearly, the overall thermoelectric efficiency is well reproduced in the larger-scale samples, which makes them promising for the fabrication of a thermoelectric module for low-temperature applications.

\begin{figure*}
\centering
% \includegraphies[width=0.9\textwidth]{Images/Voc_DT.png}
\includegraphics[width=0.9\textwidth]{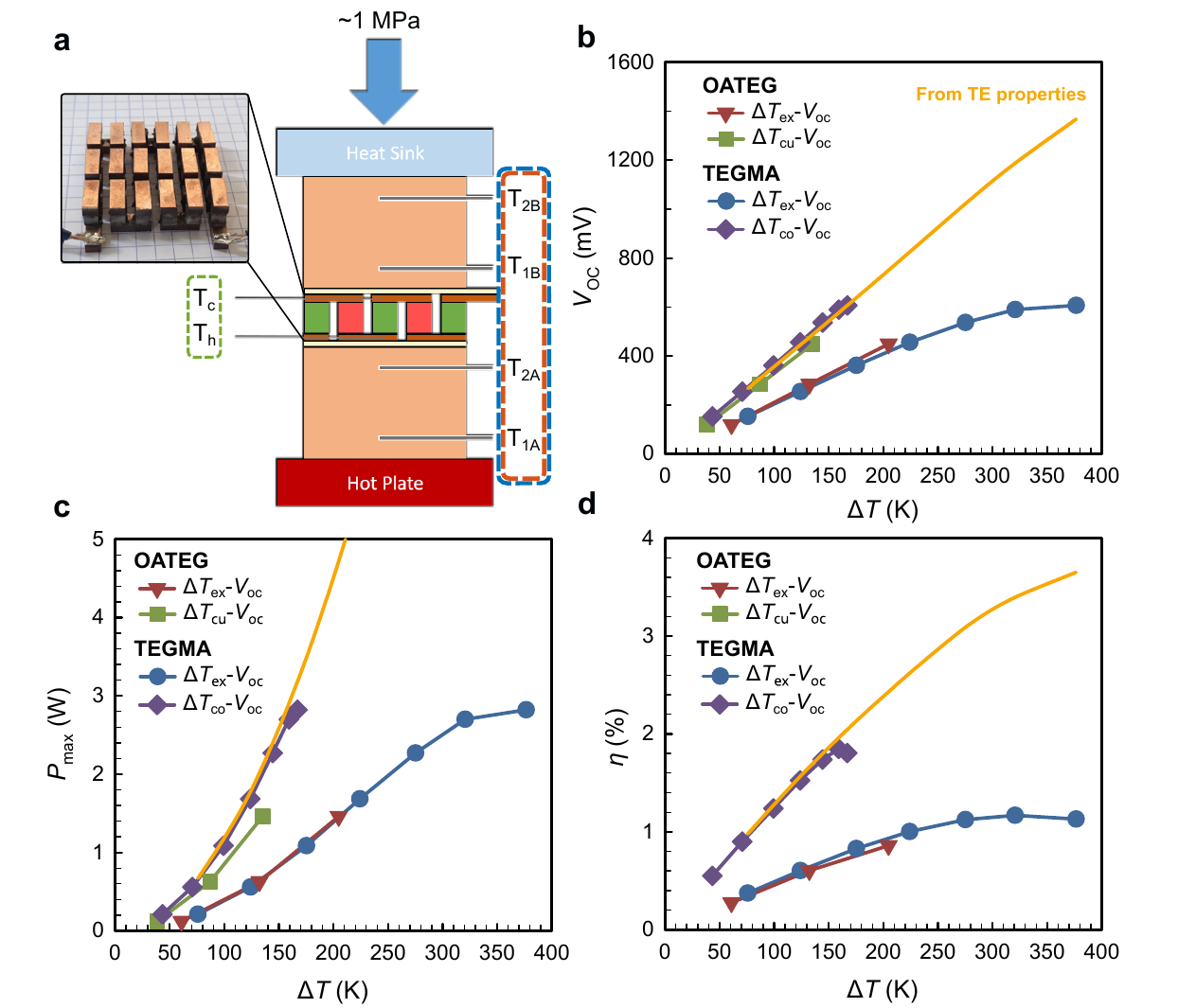}
\caption{\textbf{Performance parameters of the 36-leg \ce{Fe2VAl}-based module} \textbf{a}, Schematic view of the module characterization setup showing the different methods used to estimate the temperature difference. The inset shows a picture of the fabricated thermoelectric module. \textbf{b}, Open circuit voltage as a function of the temperature difference: HFM extrapolation (on \textsc{TEGMA} in blue, on \textsc{OATEG} in red), by direct contact on \textsc{OATEG} (in green), computed using \autoref{equ:DT_co} (in purple). \textbf{c}, Maximum electrical power as a function of temperature difference. \textbf{d}, Conversion efficiency as a function of the temperature difference.}
\label{fig:Voc_DT}
\end{figure*}
\section{Fabrication and characterization of a thermoelectric module}
\label{TEM}

To fabricate the thermoelectric module (TEM), 4$\times$4$\times$4 mm$^3$ legs were diced by electrical discharge machining from the hot-pressed disks of the $n$- and $p$-type materials. Two thermoelectric modules were assembled by direct brazing on copper connectors (2 and 36 legs, respectively) using an assembling rig inspired by the work of Fabian-Mijangos \textit{et al.} \citep{fabian2017enhanced}. The brazing cycle was conducted by induction heating under vacuum ($<10^{-4}$\,mbar) at 1073\,K for 10 minutes with a heating rate of 100\,K.min$^{-1}$ and natural cooling. This process is depicted in the Supporting Information (Fig. S2).
% \autoref{fig:assembly}.
%
% \begin{figure*}
% \centering
% \includegraphics[width=0.9\textwidth]{Images/Assembly_Setup_V1.png}
% \caption{\textbf{Brazing setup for TE module assembly} }
% \label{fig:assembly}
% \end{figure*}
%
%

Contact resistances were characterised on a two-leg module using a custom voltage scanning probe apparatus with a 200\,µm resolution. The result of the resistance scan over the whole couple is shown in \autoref{fig:rScan}. %The electrical resistivities measured from the slopes in the thermoelectric legs are within 5\,\% of the materials properties depicted in Fig. SI?, ensuring that the brazing steps did not significantly alter the TE material.
The exact contact resistances could not be determined since resistance drops are not visible at the interfaces. Nevertheless, given the resolution of the scanning apparatus, it can be stated that contact resistances are low ($<6 \mu\Omega$\,cm$^2$). Notably, this is one order of magnitude lower than values reported in previous work on brazing of \ce{Fe2VAl}-based modules \citep{roy2019global}, which can mainly be attributed to the optimization of the brazing cycle by improving the vacuum conditions and increasing the brazing temperature.
\begin{figure*}
\centering
\includegraphics[width=0.85\textwidth]{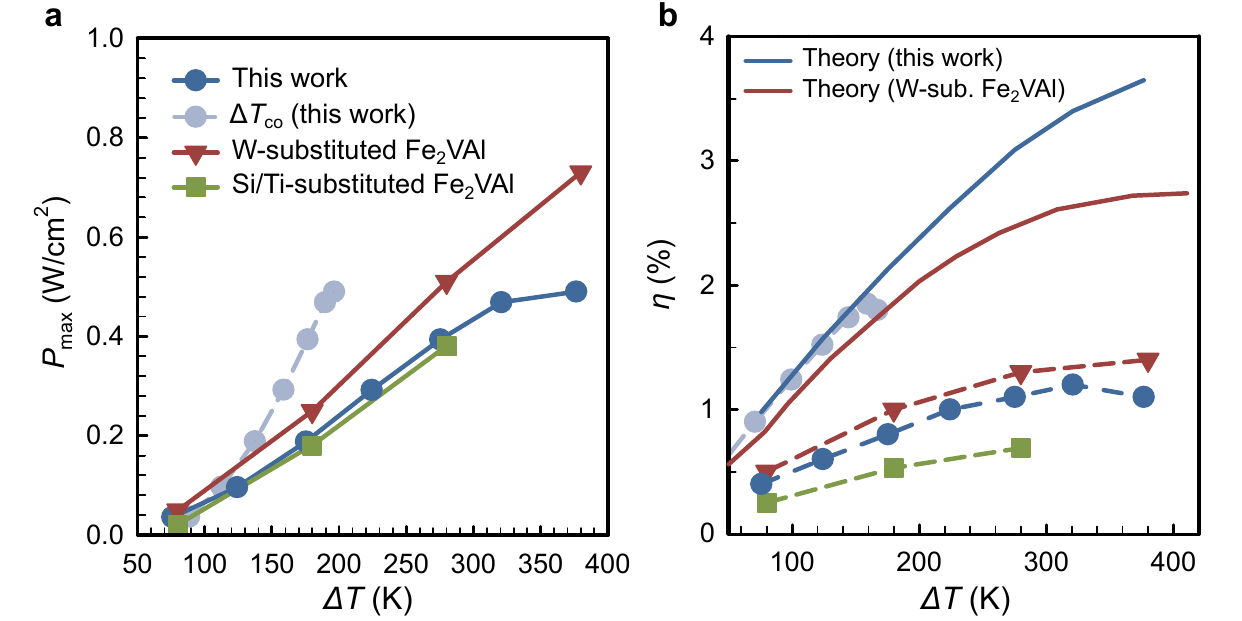}
\caption{\textbf{Comparison of \ce{Fe2VAl}-based thermoelectric modules.}
\textbf{a}, Temperature-dependent power density of the present module compared with reports by Mikami $et~al.$ \citep{mikami2009development, mikami2014evaluation}. The comparison depends on the method used to determine $\Delta T$. 
\textbf{b}, Temperature-dependent conversion efficiencies of the same modules. When determined using $\Delta T_\text{co}$, the present module exhibits the highest efficiencies reported to date for \ce{Fe2VAl}-based thermoelectric modules.}
\label{fig:perf_DT}
\end{figure*}
The 36-leg module was fully characterised regarding its thermoelectric properties utilizing two different measurement setups operated by different research groups: the \textsc{TEGMA} and the \textsc{OATEG} setups.
\begin{itemize}
    \item The measurements with the \textsc{TEGMA} setup, previously described by Ziolkowski \textit{et al.}, were conducted under vacuum \citep{ziolkowski2021heat}. Heat flow was measured using copper-based heat flow meters (HFM). The module was coupled to the measurement section of the TEGMA via graphite foils and ceramic substrates to improve thermal coupling and to prevent short circuits. To determine the temperatures at the module, the temperature profiles measured within the heat flow meters were extrapolated to the module coupling interfaces. The electric current was varied in multiple steps between open-circuit and short-circuit conditions at each temperature level. To correct for the Peltier effect on the module interface temperatures, the so-called Rapid Steady State (RSS) method was applied \citep{ziolkowski2021validation}. In this method, the current through the module is briefly interrupted to measure the decreasing open-circuit voltage caused by the reduced temperature difference due to the Peltier effect, which is then taken into account when evaluating the internal electrical resistance and electrical output power of the module.
    \item The \textsc{OATEG} is an open-air setup, using a copper HFM on the cold side for heat flow determination. The module was coupled to HFMs using graphite foil insulated with a sprayed layer of boron nitride. Two methods are available for the determination of the temperature accross the module. The first one is similar to the \textsc{TEGMA} setup by using extrapolation from HFM temperature profiles. The second approach relies on easy access to the thermoelectric module connectors, which allows direct temperature measurements on the hot and cold sides of the thermoelectric legs (see \autoref{fig:Voc_DT}). This second configuration allows neglecting the thermal contact resistances between the module and the HFM. However, it limits the hot-side temperature to 523\,K due to the sheath material of the thermocouples. At each temperature level, the electric current was varied in multiple steps between open-circuit and short-circuit conditions. No correction for the Peltier effect was applied, as it was assumed to be negligible owing to the moderate $zT$ of the materials investigated in this work.
\end{itemize}

All measurements were performed under a mechanical pressure of 1 MPa.
%
% \autoref{fig:I_curve}. The maximum measured power and conversion efficiency reach 2.8 W and 1.2\% for a temperature difference of 375 K. Relative expanded uncertainties for power and efficiency are around 1\% and 10\% respectively.
%
% \begin{figure*}
% \centering
% \includegraphics[width=0.9\textwidth]{Images/I_curve_V2.png}
% \caption{\textbf{36 legs module performances as a function of the load current for different temperature differences:} (a) voltage, (b) electrical power, (c) cold side heat flow and (d) conversion efficiency.}
% \label{fig:I_curve}
% \end{figure*}
\autoref{fig:Voc_DT}\,a shows a schematic of the measurement setups described above. While the setups differ in some details, such as the available temperature sensor locations, the schematic applies to both configurations. Notably, the measurement points directly on the TEM (marked in green on the left) are only available when using the \textsc{OATEG} setup.
\autoref{fig:Voc_DT}\,b shows the evolution of the open-circuit voltage as a function of temperature difference. Three different methods are used to estimate the effective temperature difference on the module: extrapolation in HFMs ($\Delta T_\text{ex}$), direct measurement on Cu connectors ($\Delta T_\text{cu}$) and computation from measured $V_\text{oc}$ and materials properties ($\Delta T_\text{co}$).
This last estimation is computed similarly to the work of Mikami \textit{et al.} \citep{mikami2014evaluation}:
\begin{equation}
\Delta T_\text{co}=\frac{V_\text{oc}}{N({S_p-S_n})}\;,
\label{equ:DT_co}
\end{equation}
where $V_\text{oc}$ is the measured open-circuit voltage, $N$ is the number of couples and $S_p$ ($S_n$) the Seebeck coefficient of the $p$-type ($n$-type) material.
The power generation performance as a function of electrical current for the 36-legs module was characterized over several heating–cooling cycles with a cold-side temperature of 298\,K using the \textsc{TEGMA} setup (see details in the Supporting Information (Fig. S3)). The average performance over three cycles is presented in \autoref{fig:Voc_DT}\,c and compared to \textsc{OATEG} measurements where only one cycle was performed. 
Direct measurements provide the closest agreement with the expected behavior predicted from the material properties, suggesting the presence of significant thermal contact resistances at the interface between the TEM and the HFMs. As confirmed by the evaluation of electrical power and efficiency shown in \autoref{fig:Voc_DT},c,d, the effective temperature difference is more accurately represented by $\Delta T_\text{cu}$ and $\Delta T_\text{co}$.
%
% \begin{figure*}
% \centering
% \includegraphics[width=0.9\textwidth]{Images/Perf_DT_V2.png}
% \caption{(a) Maximum electrical power as a function of temperature difference. (b) Conversion efficiency as a function of the temperature difference.}
% \label{fig:Perf_DT_V2}
% \end{figure*}

Finally, the performance of this 36-leg module is compared in \autoref{fig:perf_DT} with previously published \ce{Fe2VAl}-based TEGs \citep{mikami2009development, mikami2014evaluation}. Comparing power densities and efficiencies derived from $\Delta T_\text{co}$ (obtained from the measured $V_\text{oc}$), it is straightforward that the present module achieves the highest power density and efficiency among full-Heusler devices reported to date. 
However, for the measured values, the comparison is less direct, since Mikami \textit{et al.} used different measurement setups for determining $\Delta T$. \citep{Takazawa2006, mikami2009development, mikami2014evaluation}. 

When comparing the measurement results directly, the present module performs slightly worse than the W-based module described by Mikami \textit{et al.}, owing to the significantly larger discrepancy between measured $\Delta T$ and calculated $\Delta T_\text{co}$ in the present measurements. 
This difference may arise from variations in heat coupling quality or from different methods used to determine the temperature gradient. This observation highlights the importance of optimal thermal coupling of the module in TEG systems, particularly when using materials with relatively high thermal conductivity, such as \ce{Fe2VAl}. Notably, all reported \ce{Fe2VAl}-based TEGs (including the present study) were realized without relying on nanostructuring to enhance thermoelectric properties, a strategy that has recently shown considerable potential for this class of materials \citep{fukuta2022improving, 
garmroudi2025decoupled}. This indicates substantial potential for further performance improvements through microstructural optimization of the materials and reduction of thermal contact resistances during module integration.

\section{Conclusion}
In summary, the thermoelectric performance of a $n$-type \ce{Fe2VAl}-based full-Heusler compound were optimized through high-entropy
engineering relying on a combination of Al/V self-substitution and heavy-element co-doping on all crystallographic lattice sites. This approach yielded an average dimensionless figure of merit of 0.3 in the 300–500\,K range—the highest reported for non-nanostructured \ce{Fe2VAl}-based full-Heusler materials to date. The robustness of the strategy was demonstrated by scaling up the material synthesis via hot pressing, producing bulk samples that retained the favorable thermoelectric properties. Using these optimized materials, a 36-leg thermoelectric module (4$\times$4\,cm$^2$) was fabricated, achieving a maximum output power of 2.8\,W and a record-high conversion efficiency of 1.2\,\% for \ce{Fe2VAl}-based modules at a temperature difference of 375\,K. The present results surpass previous full-Heusler modules in terms of power density and efficiency, highlighting the potential of \ce{Fe2VAl}-based compounds as scalable, sustainable, and high-performance thermoelectric materials. 

\section{Materials and Methods}
\label{methods}
Materials were synthesized by melting the raw elements via induction heating in a water-cooled copper cold boat under inert Ar atmosphere. The as-cast materials were cut into rectangular bar-shaped samples with typical dimensions of $1.5\times 1.5 \times 10\,$mm using a high-speed cutting machine (Accutom) from the company Struers. Electrical transport measurements from 300--860\,K were performed by using a commercially available standard setup (ZEM-3 from ULVAC-RIKO). The thermal conductivity was measured with a light-flash diffusivity technique using a commercially available device (LFA-500 from Linseis).

For the up-scale synthesis of the material to be used as legs in a module, we found that induction-melted samples suffer from cracks due to the mechanical stress induced by rapid solidification from the melt. Thus, we pursued an alternative synthesis route. The melted material was ground into a fine powder using a Retsch MM 400 vibrational ball mill with a 25 ml tungsten carbide (WC) container and a single WC ball with a diameter of 15 mm. The ball milling time and frequency wer set to 10 min and 30 Hz, respectively. The obtained powder was consolidated using a hot press as described in section III. Measurements were performed as described above.

\section{Acknowledgement}
The authors acknowledge the TU Wien X-Ray Center (XRC) for the measurement time for the structural analysis of our samples. The authors acknowledge the UCLouvain LACAMI Platform for the support in module assembly and characterization.
M.\,P. and E.\,B. are grateful for  financial support by JST (Japan) in terms of the project ``MIRAI''.  F.\,G. acknowledges a
Director’s Postdoctoral Fellowship through the
Laboratory and Directed Research \& Development (LDRD) program. G.\,R. and P.\,J. acknowledge the financial support of the Walloon Region through the MultiThermEx project (Win2Wal n° 2010176).

%\bibliography{TEG_paper}
%\bibliographystyle{apsrev4-2}


\end{document}